\documentclass{iopart}

\usepackage{epsfig}
\usepackage{psfig}
\newcommand{\Z}{{\mathcal Z}}
\newcommand{\tZ}{{\tilde {\mathcal Z}}}
\newcommand{\Ze}{{\mathcal Z}_{\rm ex}}
\newcommand{\tw}{{\tilde w}}
\newcommand{\tN}{{\tilde N_0}}
\newcommand{\mN}{N_{\rm 0,max}}
\newcommand{\meps}{\epsilon_{\rm 0,max}}
\newcommand{\kb}{k_{\rm B}}
\newcommand{\mub}{\mu_{\rm B}}
\newcommand{\ef}{\epsilon_{\rm F}}

\begin{document}

\jl{1}
\title[Fermi condensation]{Condensation in ideal Fermi gases}

\author{Drago\c s-Victor Anghel \footnote{E-mail: dragos@fys.uio.no}} 

\address{University of Oslo, Department of Physics, P.O.Box 1048 
Blindern, N-0316 Oslo, Norway}

\begin{abstract} 
I investigate the possibility of condensation in ideal Fermi systems 
of general single particle density of states. 
For this I calculate the probability $w_{N_0}$ of 
having exactly $N_0$ particles in the condensate and analyze its 
maxima. The existence of such maxima at macroscopic values of 
$N_0$ indicates a condensate. An interesting situation occurs for 
example in 1D systems, where $w_{N_0}$ may have two maxima. One is at 
$N_0=0$ and another one may exist at finite $N_0$ (for temperatures 
bellow a certain condensation temperature). This suggests the 
existence of a first order phase transition. 
The calculation of $w_{N_0}$ allows for the exploration of 
ensemble equivalence of Fermi systems from a new perspective. 
\end{abstract}

\pacno{05.30.-d, 05.30.Fk, 05.30.Ch}
\submitto{\JPA}

\maketitle

\section{Introduction}

Huge progress have been made in the last decade in cooling trapped Fermi 
or Bose gases down to the Fermi degeneracy temperature or to the 
onset of the Bose-Einstein condensation, respectively (see for example 
Refs. \cite{justif,jin1,jin2,pitaevskii} and citations therein). 
Tuning experimental parameters, like the trap frequency or the 
magnetic fields applied, one can change the strength of the 
interaction between the particles and the density of the gas. 
This freedom, which allows for more accurate comparison between theory 
and experiment, led to a burst of scientific research in the field. 
The quantum effects in these gases (fermionic degeneracy and 
Bose-Einstein condensation) are usually observed at temperatures around or 
below 1 $\mu$K. 
Such temperatures can only be achieved by evaporative cooling \cite{evc}, 
but the time scale for the cooling process depends strongly on the 
elastic collision rate of the particles, which determines the thermalization 
rate \cite{Jin}. 
In the case of fermions, at temperatures below the Fermi 
temperature, the scattering probability falls dramatically due to the Pauli
blocking mechanism, which makes cooling very slow. 
In contrast to this, in the case of bosons, the presence of the 
Bose-Einstein condensate enhances the cooling rate due to the 
assisted scattering of hot particles out of the trap, and the cooling 
process can continue until almost all the gas is condensed. 
In analogy to the bosonic case, the existence of a condensate in the 
Fermi system would create the possibility of producing a highly degenerate 
Fermi gas by removing (evaporating) the particles above the condensate. 
Moreover, the Fermi condensate might have the same role of enhancing 
the cooling rate, as the Bose-Einstein condensate. 
It is well known that a Fermi gas with BCS interaction undergoes a 
phase transition to a condensed, superconducting phase \cite{BCS}. 
In this paper I shall show that a condensate may exist even in ideal 
Fermi gases. 

\begin{figure}[t]
\begin{center}
%\fbox{
\unitlength1mm\begin{picture}(20,60)(0,0)
\put(0,0){\psfig{file=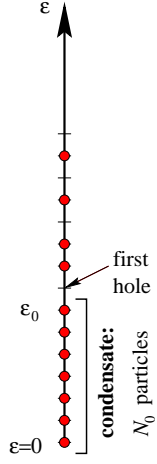,width=20mm}}
\end{picture}
%}
\caption{Single particle energy axis. Single-particle states are 
numbered from $0$ to $\infty$, starting from the lowest in energy 
and going up-wards.}
\label{fermi}
\end{center}
\end{figure}

At low temperatures, there will be a number $N_0$ of fermions that 
occupy the lowest single-particle states. If I number the single 
particle states from $0$ to $\infty$, starting from lowest energy level 
and going up-wards, then the first unoccupied state is numbered 
$N_0+1$. (The order of the degenerate levels is not important for 
macroscopic systems.)
If $N_0$ is a macroscopic number, then I will say that {\em it forms a 
condensate} (see figure \ref{fermi}). 

Usually, at low temperatures the particles below the Fermi level are 
said to be on the ``Fermi ground state'' (see for example Refs. 
\cite{tran}). The only distinctive physical property of the Fermi energy 
is that at (very) low temperatures, the particle and hole populations 
are symmetric with respect to it. 
Due to the analogy between the Bose-Einstein condensate 
(see Ref. \cite{167673}) and the Fermi condensate I suggest that the 
Fermi condensate may be better suited for the role of Fermi ground state. 

In Appendix B of Ref. \cite{start} I showed that a condensate forms 
in an interacting system (with constant density of states), with 
general microscopic exclusion statistics properties \cite{haldane}. 
At the condensation temperature the system undergoes a first order phase 
transition. In \cite{167673} I showed how the condensation 
in the corresponding noninteracting system changes the character of this 
phase transition.

\section{Condensation}

Let me consider a Fermi system at temperature $T$ and chemical potential 
$\mu$. Its grandcanonical partition function is 
\begin{equation}
\Z = \sum_m\exp\left[-\beta(U_m-\mu N_m) \right] , \label{defZ}
\end{equation}
where the sum is taken over all the microstates $m$, of internal energy 
$U_m$ and particle number $N_m$; as usual, $\beta\equiv (\kb T)^{-1}$. 
If the system is ideal, then $N_m=\sum_{i=0}^\infty n_{i,m}$ and 
$U_m=\sum_{i=0}^\infty n_{i,m}\epsilon_i$, where by $\epsilon_i$ I denote 
the energy of the single particle states. I take $\epsilon_0=0$ and 
$\epsilon_i\le\epsilon_{i+1}$, for any $i\ge 0$, as mentioned in the 
Introduction section. On each single particle state 
there are $n_{i,m}=0$ or 1 particles. 
Introducing the expressions for $U_m$ and $N_m$ into equation (\ref{defZ}), 
I can rewrite $\Z$ in the well-known form, 
\begin{equation} \label{defZ1}
\Z = \prod_{i=0}^\infty \sum_{n_i=0}^1 \exp[-\beta(\epsilon_i-\mu)n_i]  = 
\prod_{i=0}^\infty \left\{1+\exp[-\beta(\epsilon_i-\mu)]\right\}\, .
\end{equation}
The logarithm of $\Z$ is related to the grandcanonical thermodynamic 
potential, 
\begin{equation} \label{Ztot}
-\beta\Omega\equiv\log\Z=\sum_{i=0}^\infty \log\left[1+\e^{\beta(\mu-\epsilon_i)} \right] \, .
\end{equation}
The probability of the microstate $m$ is 
$w_m=\Z^{-1}\exp\left[-\beta(U_m-\mu N_m) \right]$. 

I assume that for particles in an arbitrary trap and arbitrary 
number of dimensions, the density of states (DOS) has the expression 
$\sigma(\epsilon)\equiv C\epsilon^s$, where $C$ and $s$ are constants 
(I take $s>-1$). 
For example if particles of mass $m$ are in a $d$-dimensional box of 
volume $V$, then $C=V(2\pi)^{-d/2}\Gamma^{-1}(d/2+1)d(m/\hbar^2)^{d/2}$ 
and $s=d/2-1$. 
I shall use the same procedure as in Ref. \cite{start}, to calculate the 
number of condensed particles. For this, let me calculate the probability 
that the lowest $N_0+1$ states are completely occupied, i.e. 
$n_{0\le i \le N_0}=1$ and $n_{i>N_0}$ may be either 0 or 1. 
The probability of such a configuration is 
\begin{eqnarray} 
\fl
\tw_{N_0+1} &=& \Z^{-1}\cdot\exp\left[-\beta\cdot\sum_{i=0}^{N_0}\epsilon_i
-\beta\mu(N_0+1)\right]\cdot\prod_{i=N_0+1}^\infty \sum_{n_i=0}^1 
\exp[-\beta(\epsilon_i-\mu)n_i] \nonumber \\
\fl
&=& \Z^{-1}\cdot\exp\left[-\beta\cdot\sum_{i=0}^{N_0}\epsilon_i
-\beta\mu(N_0+1)\right]\cdot\prod_{i=N_0+1}^\infty 
\left\{1+\exp[-\beta(\epsilon_i-\mu)]\right\} \, . \label{twd}
\end{eqnarray}
Using the expression for $\sigma$ to transform the summation 
$\sum_{i=0}^{N_0}\epsilon_i$ into an integral and denoting 
$\Ze(N_0,\beta,\beta\mu)\equiv\prod_{i=N_0+1}^\infty \left\{1+\exp[-\beta(\epsilon_i-\mu)]\right\}$, equation (\ref{twd}) becomes 
\begin{equation} \label{tw}
\fl
\tw_{N_0+1} = \exp\left[ -\beta C\frac{\epsilon_0^{s+2}}{s+2}
+\beta\mu C\frac{\epsilon_0^{s+1}}{s+1} \right] \cdot 
\frac{\Ze(N_0,\beta,\beta\mu)}{\Z}\equiv \frac{\tZ_{N_0}}{\Z} \, ,
\end{equation}
where $\epsilon_0$ is given by the equation 
$N_0=C\epsilon_0^{s+1}/(s+1)$ (is the level up to which all the states 
are occupied by $N_0+1$ particles), while $\Ze(N_0,\beta,\beta\mu)$ is 
the partition function of the particles above the level $\epsilon_0$. 
Note that $\sum_{N_0} \tw_{N_0} \ne 1$ and $\tZ_{N_0=0}\equiv\Z$.
$\tw_{N_0+1}$ is the probability that {\em at least} $N_0+1$ particles are 
condensed. 

The quantity of direct relevance here is the probability to have 
{\em exactly} $N_0$ particle condensed. Such a configuration is 
obtained by removing the particle from the level 
$\epsilon_{N_0}=\epsilon_0$. Doing this in equation (\ref{twd}) or 
(\ref{tw}), I obtain the probability
\begin{equation} \label{wN0}
\fl
w_{N_0} = \exp\left[ -\beta \left(C\frac{\epsilon_0^{s+2}}{s+2}-
\epsilon_0\right)+\beta\mu \left(C\frac{\epsilon_0^{s+1}}{s+1} -1
\right) \right] \cdot 
\frac{\Ze(N_0,\beta,\beta\mu)}{\Z}\equiv \frac{\Z_{N_0}}{\Z} \, .
\end{equation}
Note that 
$w_{\epsilon_0}\equiv\sigma(\epsilon_0)w_{N_0(\epsilon_0)}$ 
is the probability density along the energy axis. Nevertheless, in 
what follows I shall work with $w_{N_0}$. 
If I denote the average of $N_0$ by 
$\tN\equiv\sum_{N_0} N_0 w_{N_0}$, then $\tN$ is macroscopic 
if (but, in principle, not only if) $w_{N_0}$ has a maximum at $\mN$, so that 
in the thermodynamic limit $\mN/N > 0$, or $\meps > 0$ 
[where $\mN\equiv C\meps^{s+1}/(s+1)$]. 
Since $\Z$ does not depend on $N_0$, the maximum of $w_{N_0}$ 
may be found by solving the equation 
$\partial \log \Z_{N_0}/\partial N_0 = 0$, or 
$\partial \log \Z_{N_0}/\partial \epsilon_0 = 0$. 
First I calculate $\log\Z_{N_0}$ from equation (\ref{wN0}): 
\begin{eqnarray} 
\log\Z_{N_0} &=& \left[ -\beta \left(C\frac{\epsilon_0^{s+2}}{s+2}-
\epsilon_0\right)+\beta\mu \left(C\frac{\epsilon_0^{s+1}}{s+1} -1
\right) \right] \nonumber \\
&&+ C\int_{\epsilon_0}^\infty \rmd\epsilon\, \epsilon^s 
\log\left[1+\e^{-\beta(\epsilon-\mu)} \right] \, ,\label{logZN0}
\end{eqnarray}
where the integral represents $\log \Ze(N_0,\beta,\beta\mu)$. 
The derivative of equation (\ref{logZN0}) gives 
\begin{eqnarray}
\frac{\partial\log \Z_{N_0}}{\partial \epsilon_0} &=& 
-C\beta \epsilon_0^{s+1}+\beta+C\beta\mu \epsilon_0^{s}- 
C\epsilon_0^s\log\left[1+\e^{-\beta(\epsilon_0-\mu)} \right] \nonumber \\
&=& -C\epsilon_0^s\left\{\log\left[1+\e^{\beta(\epsilon_0-\mu)} \right]
- \frac{\beta}{C\epsilon_0^s}\right\} \, , \label{HA!}
\end{eqnarray}
or, without assuming anything about $\sigma$, 
$\partial\log\Z_{N_0}/\partial\epsilon_0=-\sigma(\epsilon_0)\left\{\log\left[1+\e^{\beta(\epsilon_0-\mu)} \right]-[\sigma(\epsilon_0)\kb T]^{-1}\right\}$. 

The interpretation of equation (\ref{HA!}) is simple. 
First note that $\partial \log \Z_{N_0}/\partial \epsilon_0$ is continuous 
on the interval $(0,\infty)$. If $\epsilon_0\approx\mu$, then 
$\log\left[1+\e^{\beta(\epsilon_0-\mu)} \right]$ is of the order 1, 
so $\partial\log\Z_{N_0}/\partial\epsilon_0 < 0$, since in the 
thermodynamic limit $[\sigma(\epsilon_0)\kb T]^{-1}\ll 1$.
Moreover, for $\beta(\epsilon_0-\mu)\gg 1$, 
\[
\frac{\partial\log \Z_{N_0}}{\partial \epsilon_0} \approx 
-C\epsilon_0^s\beta\left\{\epsilon_0-\mu- \frac{1}{C\epsilon_0^s}\right\}\, ,
\]
and because $s>-1$, $\partial\log\Z_{N_0}/\partial\epsilon_0$
becomes negative for large enough $\epsilon_0$, as one would expect 
(the probability to have an infinite number of particles condensed is 
zero). 
To study in more detail the existence and the number of solutions, 
let me divide the problem into three cases: (1) $s=0$, (2) $s<0$, 
and (3) $s>0$. 

%***************************************************
\paragraph{Case (1)} is the simplest and corresponds, for example, to a Fermi 
gas in a two-dimensional (2D) box. This case may be analyzed in 
connection with the Bose-Fermi thermodynamic equivalence in (2D), which 
was outlined in Section 2.2 of Ref. \cite{start}. 
The properties of ideal 2D Bose gases have been studied extensively in the 
past (see for example \cite{holthaus&Co} and citations therein).
For $s=0$  equation (\ref{HA!}) becomes:
\begin{eqnarray}
\left.\frac{\partial\log \Z_{N_0}}{\partial \epsilon_0}\right|_{s=0} 
&=& -C\left\{\log\left[1+\e^{\beta(\epsilon_0-\mu)} \right]
- \frac{1}{C \kb T}\right\} \, . \label{HA!0}
\end{eqnarray}
If I take the thermodynamic limit simply as $(C\kb T)^{-1}=0$, then 
$\left.\partial\log\Z_{N_0}/\partial\epsilon_0\right|_{s=0}<0$ for 
all $\epsilon_0$ and $T$, so there is no condensate.
On the other hand, note that for $\beta(\epsilon_0-\mu)\ll-1$, 
$\log\left[1+\e^{\beta(\epsilon_0-\mu)}\right]\approx\e^{\beta(\epsilon_0-\mu)}$ and, because of the exponential dependence on $\beta(\epsilon_0-\mu)$, 
the term in curly brackets in equation (\ref{HA!0}) may become 
negative even for a macroscopic system. Since $\log\left[1+\e^{\beta(\epsilon_0-\mu)} \right]$ is monotonically increasing with $\epsilon_0$, while 
$(C \kb T)^{-1}$ is a constant, equation (\ref{HA!0}) has maximum one 
solution, and this must be for $0\le\epsilon_0<\mu$. I define the condensation 
temperature $T_{\rm c,2D}$ by the equation 
%\begin{equation} \label{condT}
$(\partial\log\Z_{N_0}/\partial\epsilon_0)_{T=T_{\rm c,2D},\epsilon_0=0}=0$.
%\end{equation}
In the approximation $\beta\mu\gg 1$ this condition gives:
\begin{equation} \label{c2D}
\frac{\mu}{\kb T_{\rm c,2D}}=\log [(\kb T_{\rm c,2D}) C].
\end{equation}
For any $T>T_{\rm c,2D}$, equation (\ref{HA!0}) has no solution, while 
for $T\le T_{\rm c,2D}$ it has one and only one solution for 
$\epsilon_0\ge 0$, so the system is condensed. 
To see if equation (\ref{c2D}) has any relevance for macroscopic systems, 
let me consider a 2D gas of electrons of Fermi energy 
$\ef^{\rm Al}=11.7$ eV (Fermi energy of electrons in Al). If I take 
the area of the 2D gas to be $S=1\ {\rm m}^2$ (surely a macroscopic system) 
and $\mu=\ef^{\rm Al}$, then 
\begin{equation} \label{cAl}
T_{\rm c,2D}^{\rm Al} \approx 3263 \ {\rm K}\, . 
\end{equation}
At room temperature $T_{\rm r}=300$ K, keeping $\mu=\ef^{\rm Al}$, I obtain 
%\begin{equation} \label{meps2D}
$\meps^{\rm Al} \approx 10.7$  eV,
%\end{equation}
only about 1 eV smaller than $\mu$. If I take $\tN\approx \mN$, then 
the condensate fraction ($\tN/N$) is 
$\meps^{\rm Al}/\ef^{\rm Al}\approx 0.91$. 

The condensation energy scales roughly with the chemical potential. 
For another example I take heavily doped silicon with the 
doping concentration of $n=4\times 10^{25}\ {\rm m}^{-3}$ \cite{savin}. 
For this concentration, the Fermi energy is $\ef^{\rm Si}\approx 0.04$ eV, 
which gives a condensation temperature of about 14 K. In this case, 
the condensate fraction at, say 100 mK (roughly the working temperature for 
the microcoolers or microbolometers), is over 99\%. 

Two systems of equal number of particles are called {\em thermodynamically 
equivalent} if they have the same entropies as function of temperature. 
All 2D ideal gases of equal DOS are thermodynamically equivalent, 
irrespective of their microscopic exclusion statistics 
\cite{may,lee1,lee2,lee3,lee4,apostol,viefers,isakov2,start}, and 
this is due to the similarity between their excitation spectra \cite{start}. 
If I put in correspondence the Fermi system under investigation with 
the equivalent Bose system, and I denote by $\mub$ the chemical potential 
of the later, then $\mub=\mu-\ef$ (where $\ef$ is the Fermi energy). 
The number $N_0$ of 
fermions in the condensate is equal to the number of bosons on the 
ground state, say $N_{\rm B,0}$ \cite{start}. Moreover, in the 
canonical ensemble the condensate fluctuations are the same for 
the two, equivalent, Bose and Fermi gases. 

\paragraph{Case (2)} is maybe the most interesting.
For $s<0$, $\partial \log \Z_{N_0}/\partial \epsilon_0 <0$ in both limits. 
$\epsilon_0\to 0$ and $\epsilon_0\to \infty$, so eventual solutions 
of the equation $\partial \log \Z_{N_0}/\partial \epsilon_0=0$ come 
in pairs. 
To see that there is only one pair of solution note that that the 
derivative $\partial\log\left[1+\e^{\beta(\epsilon_0-\mu)} \right]/\partial\epsilon_0=\beta/[\exp(\beta\mu-\beta\epsilon_0)+1]$ increases with $\epsilon_0$, 
so $\log\left[1+\e^{\beta(\epsilon_0-\mu)} \right]$ is a 
function concave up-wards, while 
$1/C\epsilon_0^s$ is a function concave down-wards ($-1<s<0$). Therefore 
the two curves ($\log\left[1+\e^{\beta(\epsilon_0-\mu)} \right]$ and 
$\beta/C\epsilon_0^s$) may cut each-other  either in 
exactly two points, or in zero points. 
Moreover, if I assume that they do cut and at $\epsilon_0\approx\mu$, or 
bigger, then $\kb T\cdot\{\partial\log\left[1+\e^{\beta(\epsilon_0-\mu)} \right]/\partial\epsilon_0\}=[\exp(\beta\mu-\beta\epsilon_0)+1]^{-1}$ is of the 
order one or bigger, while 
$-\kb T\cdot[\rmd (1/C\epsilon_0^s)/\rmd\epsilon_0]=-s\kb T/[C\epsilon_0^s\kb T\cdot(\epsilon_0/\kb T)]\ll 1$, in the thermodynamic limit. 
Therefore, $\partial\{\log\left[1+\e^{\beta(\epsilon_0-\mu)} \right]- \frac{\beta}{C\epsilon_0^s}\}/\partial \epsilon_0 >0$ for $\epsilon_0\approx\mu$, or 
bigger. Since $\log\left[1+\e^{\beta(\epsilon_0-\mu)} \right]- \frac{\beta}{C\epsilon_0^s}>0$, for $\epsilon_0\approx\mu$, I conclude that 
$\partial\log \Z_{N_0}/\partial \epsilon_0=0$ has no solutions for 
$\epsilon_0\in[\mu, \infty)$, no matter what is the value, or sign of 
$\mu$. If $\partial\log \Z_{N_0}/\partial \epsilon_0=0$ has solutions, then 
$\mu>0$ and the solutions are in the interval $\epsilon_0\in (0,\mu)$. 

I define the condensation temperature by the equation 
\begin{equation} \label{cond1D}
\max_{\epsilon_0}\left.\frac{\partial\log\Z_{N_0}}{\partial
\epsilon_0}\right|_{T=T_{\rm c,s<0}} = 0 \, .%\ ({\rm for\ s=-1/2}) \, .
\end{equation}
For $T>T_{\rm c,s<0}$, equation (\ref{HA!}) has no solution, so the 
system is not condensed. 
Particles on a one-dimensional (1D) interval have 
$\sigma(\epsilon)=C\epsilon^{-1/2}$, so I shall take 1D electron 
systems as examples. If I take again the Fermi energy of Al 
for electrons on an interval of length $l=1$ m, I obtain 
$T_{\rm c,1D}\approx 6242$ K. 

The interesting aspect about the $s<0$ case is that below the 
condensation temperature $\Z_{N_0}(\epsilon_0)$ has two maxima: 
one at $\epsilon_0=0$ and one at $\epsilon_0=\meps>0$. By comparing 
the areas below the peaks, one can conclude which maximum is the 
stable solution and which is the metastable one. Similar to the 
situation outlined in Appendix B of Ref. \cite{start}, this 
leads to a first order phase transition from an uncondensed, 
to a condensed phase, in 1D {\em canonical} Fermi systems. 

\paragraph{Case (3)} includes three-dimensional (3D) ideal systems, 
which correspond to $s=1/2$. 
If $s>0$, obviously  $\partial \log \Z_{N_0}/\partial \epsilon_0 >0$ 
at $\epsilon_0=0$. 
Since $\log\left[1+\e^{\beta(\epsilon_0-\mu)} \right]$ is monotonically 
increasing and $(C\kb T\epsilon_0^s)^{-1}$ is monotonically decreasing, while 
$\partial \log \Z_{N_0}/\partial \epsilon_0$ is negative at large 
$\epsilon_0$, I conclude that exists one and only one finite $\meps$, and 
therefore a finite $\mN$, at which $w_{\mN}$ attains 
its maximum. For example for one cubic meter of Al, at room 
temperature, $\meps\approx 11.2$ eV and the condensate fraction 
is about 94 \%.

An important example is taken from the experiments performed by 
D. Jin and collaborators \cite{jin1,jin2} and I shall use parameters 
from Ref.  \cite{jin2}. I describe $N=6\times 10^6$ 
atoms of $^{40}$K (I take both spin polarizations), confined in a 3D 
harmonic trap of (geometric) mean frequency $\omega/2\pi=70$ Hz 
($\hbar\omega\approx 2.9\times 10^{-11}$ eV). 
The DOS is $\sigma(\epsilon)=\epsilon^2/(\hbar\omega)^3$ and the Fermi 
energy is $\ef=(3N)^{1/3}\hbar\omega\approx 7.6\times 10^{-11}$ eV. 
With this parameters I calculate two values of $\meps$: 
$\meps(T_1=1\ {\rm \mu K})\approx 2.6\times 10^{-12}$ eV and 
$\meps(T_2=300\ {\rm nK})\approx 1.3\times 10^{-11}$ eV. Note 
that both, $\meps(T_1)$ and $\meps(T_2)$ are smaller than $\hbar\omega$, 
so there is no condensate, although the equation 
$\partial \log \Z_{N_0}/\partial \epsilon_0 =0$ has a solution. 
Eventually the possibility to obtain a highly degenerate Fermi gas is 
to agglomerate may particles in a trap and then to separate the condensate 
formed at the ``bottom'' of the trapping potential.

\section{Conclusions}

In this paper I discussed the formation of a condensate in ideal 
Fermi systems with the density of single particle states of the 
form $\sigma(\epsilon)=C\epsilon^s$ (where $C$ and $s$ are 
constants). I did this by calculating the probability $w_{N_0}$ of 
having exactly $N_0$ particles in the condensate. 
If $s>0$ (like in 3D boxes or traps), $w_{N_0}$ has a maximum 
for $N_0>0$, which indicates the formation of a condensate. The 
maximum persists at any temperature, but eventually $N_0$ becomes 
microscopical as $T$ increases. For a cubic meter of Al at room temperature, 
$N_0/N\approx 0.94$ (where $N$ is the total number of particles in 
the system). Unfortunately for the current experiments in harmonic 
traps, $N_0$ appears to be smaller than one, which means that there is 
no condensate. Maybe a high increase of particle number in the trap, without 
much concern for the temperature, would lead to the formation of 
a condensate, which may be afterwords separated by evaporating the 
uncondensed particles. 

The case $s=0$ corresponds to 2D boxes or 1D harmonic traps. This is 
equivalent to a 2D Bose gas and the results of canonical calculations 
may be interchanged down to the microscopic scale, with proper 
redefinition of single particle energies (see Ref. \cite{start}). 
After the interchange, the particles in the Fermi condensate 
become the particles onto the Bose ground state, which have been extensively 
studied \cite{holthaus&Co}.

Maybe the most interesting case is $s<0$. Here $w_{N_0}$ may have 
two maxima. One maximum is always at $N_0=0$ (no condensate), while the 
other maximum forms at finite $N_0$, for temperatures below a condensation 
temperature. The existence of the two maxima suggest a phase transition 
of order one in canonical Fermi systems and a reconsideration of the 
ensemble equivalence. In this category are included 1D Fermi systems.

\end{document}